%% file: main.tex
\documentclass[%
 reprint,
 amsmath,amssymb,
 aps,
pra,
floatfix,
]{revtex4-2}

\usepackage{xcolor}
\usepackage{graphicx}
\usepackage{dcolumn}
\usepackage{bm}

\begin{document}
\input{Tex/Title}
\input{Tex/Abstract}

\maketitle


\input{Tex/Introduction}

\input{Tex/Concept2}
\input{Tex/Modeling}

\input{Tex/Fabrication}
\input{Tex/Results}

\input{Tex/Conclusion}
\input{Tex/Acknowledgement}
\input{Tex/DataAvailibility}
\appendix
\input{Tex/AppendixA}
\input{Tex/AppendixB}


\bibliography{Bibliography}

\end{document}

%% file: Tex/Title.tex
\preprint{APS/123-QED}

\title{Superconducting Nanowire Based Surface Acoustic Wave Transduction}

\author{Jack Guida}
 \email{guida.j@northeastern.edu}
\author{Kartikey Agarwal}%
\author{Marco Colangelo}%
\author{Siddhartha Ghosh}%

\affiliation{Northeastern University, Boston, Massachusetts, USA
}%

\date{\today}

%% file: Tex/Abstract.tex
\begin{abstract}

This work presents the  demonstration of a superconducting niobium nitride (NbN) nanowire surface acoustic wave transducer, enabling a cryogenic acoustic delay line in scandium aluminum nitride (ScAlN) on silicon carbide (SiC). The superconducting nanowire slows the effective electromagnetic phase velocity along the acoustic wave propagation axis to match that of the surface acoustic wave, such that the co-propagating electrical signal continuously and coherently drives the acoustic wave along the length of the wire. The operating principle is validated through coupled full-wave electromagnetic and piezoelectric finite element method simulations and then experimentally confirmed through demonstration of a 500 $\mu$m delay line exhibiting a 125 ns time delay, with broadband transduction characterized across 0.5-20 GHz at 0.9 K, establishing superconducting nanowires as an emerging class of acoustic transducers with direct implications for cryogenic signal processing and integration with superconducting quantum circuits.
\end{abstract}

%% file: Tex/Introduction.tex
\input{Figs_Tex/Concept}

\section{\label{sec:Intro} Introduction}
The emergence of quantum computing and superconducting devices has created new demands for ultra-high performance RF components operating at cryogenic temperatures \cite{delsing_2019_2019}. Superconducting nanowire single photon detectors (SNSPDs) have demonstrated that such nanowire devices can achieve exceptional performance at superconducting operating temperatures \cite{korzh_demonstration_2020}, suggesting their utility beyond photon detection toward a broader class of cryogenic devices. Acoustic devices based on piezoelectric platforms offer a natural complement to superconducting device architectures, and are attractive in microwave signal processing due to their compact footprint, low insertion loss, and compatibility with standard semiconductor fabrication processes \cite{hackett_towards_2021}.

Surface acoustic wave (SAW) devices have found widespread use in RF filtering \cite{takaki_murata_surface_2008} and sensing applications \cite{soluch_design_1998}, and their potential application towards cryogenic components has drawn increasing interest given their ability to mediate coupling between microwave photons and phonons \cite{blesin_bidirectional_2024,schuetz_universal_2015}. Furthermore, acoustic resonators have demonstrated improvements in quality factor at cryogenic temperatures, suggesting that cryogenic operation of acoustic wave devices is not only compatible but advantageous for achieving optimal performance \cite{kramer_cryogenic_2026}.

Slow-on-fast piezoelectric platforms have gained attention following the developments in phononic integrated circuits (PnICs) utilizing SAWs \cite{guida_design_2025}. In these heterostructures, a low-acoustic-phase-velocity piezoelectric material is deposited on a high-acoustic-phase-velocity substrate, creating an acoustic impedance mismatch that confines and guides the acoustic mode, yielding devices with low insertion loss and low propagation loss. Scandium aluminum nitride (ScAlN) has emerged as a viable option in this context, offering enhanced piezoelectric coupling coefficients relative to aluminum nitride \cite{piazza_piezoelectric_2012,pinto_cmos-integrated_2022}, with stronger electromechanical coupling and a reduced acoustic phase velocity. Simultaneously, silicon carbide (SiC) serves as a high-velocity substrate, making ScAlN on SiC a promising platform for high-performance acoustic devices \cite{guida_phononic_2025}. 

In piezoelectric platforms, the transducer dictates overall device performance. Standard interdigitated transducers (IDTs) utilize alternating polarity electrodes with a lithographically defined pitch to couple an oscillating electric field to a mechanical mode through the piezoelectric effect, launching and receiving acoustic waves at a frequency set by the relationship between the acoustic phase velocity and electrode periodicity \cite{hashimoto_surface_2000}. The operating frequency of conventional IDTs is therefore fundamentally limited by the minimum feature size achievable in lithography, and the electromagnetic excitation frequency and acoustic operating frequency remain intrinsically coupled through the electrode geometry. 

Superconducting niobium nitride (NbN) nanowires exhibit zero DC resistance below their critical temperature \cite{guo_fabrication_2020}. When arranged in a meandered geometry, the nanowire supports a propagating electromagnetic wave that produces a spatially periodic field distribution analogous to the alternating electrode polarity of a standard IDT. This occurs only at specific electromagnetic resonance frequencies determined by the segment aperture. At these frequencies, the meander pitch simultaneously sets the spatial wavevector necessary to couple to a target SAW mode, while the superconducting wire geometry slows the effective electromagnetic phase velocity along the propagation axis to approximately match that of the SAW. Efficient transduction, therefore, requires simultaneous satisfaction of both conditions, and a single device geometry is capable of accessing multiple acoustic mode orders at distinct electromagnetic resonance frequencies.

While the traveling-wave transduction concept using superconducting nanowires was recently proposed~\cite{mccaughan_2025}, no experimental demonstration or numerical modeling has been reported. In this work, we present experimental validation of this transduction principle. This is supported by a multiphysics simulation framework and realized using an acoustic delay line on a ScAlN-on-SiC platform. The concept and working principle of meandered superconducting nanowires as acoustic transducers are presented in Sec.~\ref{sec:Concept}, including a physical justification by analogy to traditional IDTs. Full electromagnetic modeling of the superconducting nanowire is performed, with the resulting field distribution identity-mapped into coupled piezoelectric physics to numerically validate acoustic wave transduction in Sec.~\ref{sec:Modeling}. Experimental validation through an acoustic delay line implementation is provided in Sec.~\ref{sec:Results}, followed by concluding remarks and a discussion of future applications and scientific impact in Sec.~\ref{sec:Conclusion}.

%% file: Figs_Tex/Concept.tex
\begin{figure*}[t!]
\centering
\includegraphics[width=\linewidth]{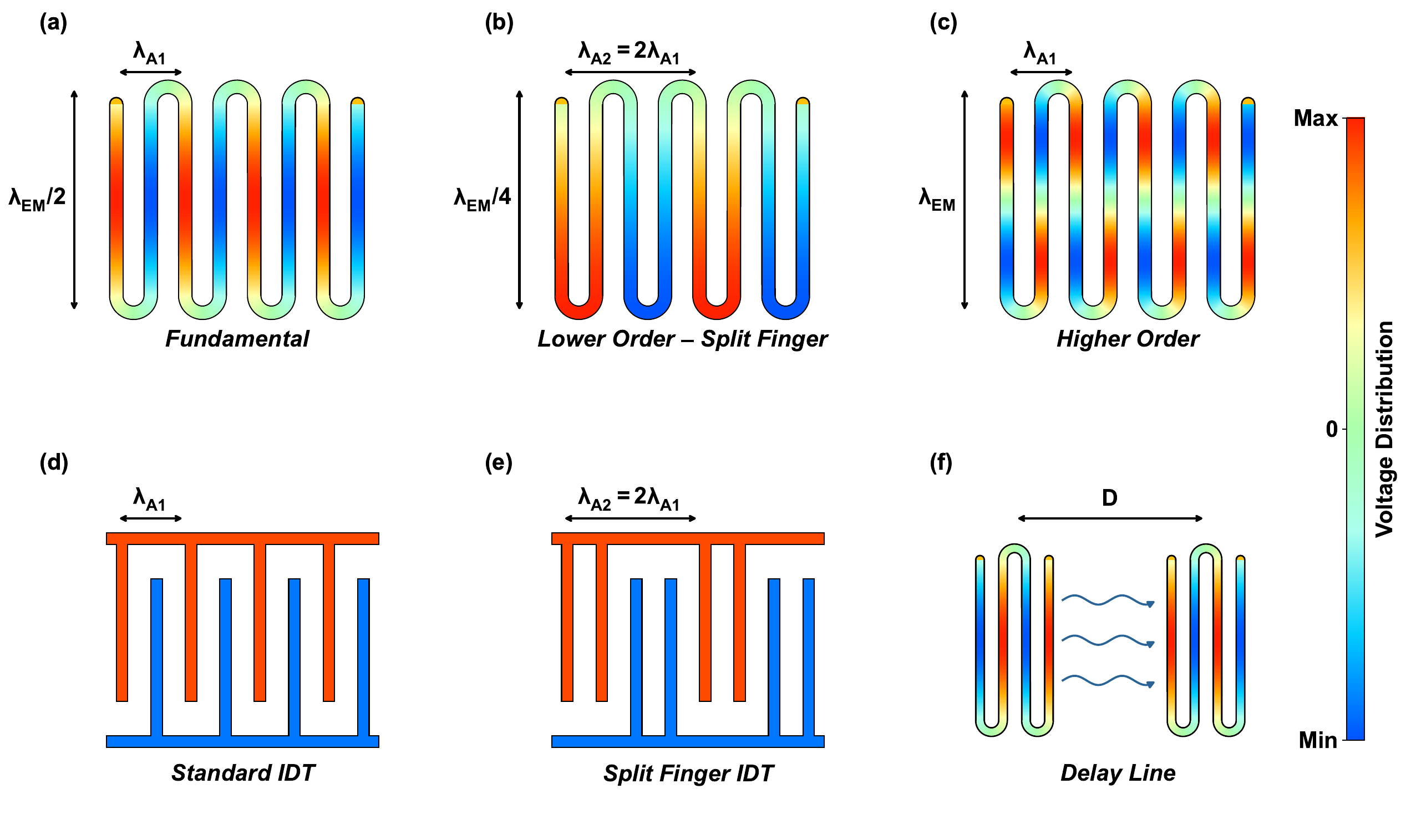}
\caption{Illustration of the meandered nanowire transducer operating modes. The spatially periodic nanowire voltage distribution mimics the alternating electrode polarity of a standard IDT. (a) Fundamental mode in which the aperture spans $\lambda_{EM}/2$, producing a single period of alternating polarity across one acoustic wavelength ($\lambda_{A1}$). (b) Lower order split finger mode, where the aperture spans $\lambda_{EM}/4$ and the acoustic wavelength doubles ($\lambda_{A2}=2\lambda_{A1}$) compared to the fundamental mode. (c) Higher order mode in which the aperture spans $\lambda_{EM}$, while exciting a higher electromagnetic harmonic. The nanowire transducers are compared directly against the (d) standard and (e) split finger IDTs. The implementation of nanowire transducers in a delay line configuration, demonstrated in this work, is shown in (f), with the blue arrows indicating SAW mode propagation, with the delay line distance indicated with D.}
\label{fig:Concept}
\end{figure*}

%% file: Tex/Concept2.tex
\section{\label{sec:Concept} Device Operating Principle}
\subsection{Meandered Nanowire as a Distributed IDT}
In traditional IDTs, periodically spaced electrodes driven at alternating polarity generate a fringing electric field that, through the piezoelectric constitutive relations, couples electromagnetic excitation to elastic acoustic modes in the underlying material. The acoustic wavelength is lithographically defined by the electrode pitch, and the operating frequency of the transducer is set by the ratio of the acoustic phase velocity to the electrode-defined wavelength ($f = v_p/\lambda_A$). The electrodes act as quasi-static capacitive elements that produce fringing fields which directly excite acoustic waves through their spatial periodicity, with the electromagnetic drive and acoustic frequencies inherently coupled at acoustic resonance.

The meandered nanowire transducer in this work operates under a fundamentally different principle than a standard IDT. Instead of discrete interdigitated electrodes, a single continuous nanowire is patterned in a serpentine meander, acting as an electromagnetic transmission line along its length. With this configuration, the contributions to the acoustic mode transduction involve the meander pitch and aperture. The meander pitch defines the acoustic wavevector ($k_A = 2\pi/\lambda_A$), selecting the wavelength of the acoustic mode that is efficiently coupled to the transducer. The meander aperture sets the input electromagnetic frequency needed for efficient electromechanical transduction. For this to occur, the electromagnetic field pattern must remain coherent with the propagating acoustic wave along the interaction length, requiring both temporal frequency matching and spatial phase matching. The first aspect is achieved when the effective electromagnetic phase velocity along the SAW propagation axis is slow enough to approximately match that of the launched acoustic mode, which is governed by the effective dielectric constant $\epsilon_{eff}$ of the nanowire stack. 

Inductance in superconductors is dominated by the inertia of the charge-carrying Cooper pairs, and is therefore known as the kinetic inductance \cite{kreidel_measuring_2024}. Since Cooper pairs have mass, accelerating or decelerating them requires energy, and hence there is inherent inductance due to their kinetic energy. Kinetic inductance is critical to calculating the effective dielectric constant, as it substantially slows down the propagating electromagnetic wave \cite{santavicca_microwave_2016}. The spatial phase matching condition is satisfied when the input electromagnetic wave produces a field distribution analogous to that of a standard IDT, where there is alternating polarity excitation in adjacent segments of the nanowire. This occurs when the aperture is half of the effective electromagnetic wavelength, $\lambda_{EM} = \lambda_o/\sqrt{\epsilon_{eff}}$. When both phase matching conditions are met, the nanowire contributes coherently to transduction over its full length, realizing a distributed unidirectional IDT analog.

Similarly, for any given device geometry, multiple electromagnetic frequencies can produce spatial field distributions with appropriate phase matching conditions, giving rise to higher and lower order modes in addition to a fundamental mode, as demonstrated in Fig.~\ref{fig:Concept}, where each conducting material shows the respective voltage distribution. Here, Fig.~\ref{fig:Concept}(a) shows the fundamental mode, where each straight nanowire segment carries alternating field polarity across adjacent sections. Halving this periodicity (doubling the electromagnetic drive frequency) yields the split-finger configuration in Fig.~\ref{fig:Concept}(b) \cite{da_cunha_network_1993}. Lastly, the higher order mode, shown in Fig.~\ref{fig:Concept}(c), spans a full electromagnetic wavelength in a single straight section. This architecture relaxes the lithographic constraints on maximum operating frequency imposed by conventional IDT electrode pitch scaling and mitigates the phonon-phonon propagation losses that dominate at higher frequencies~\cite{tabrizian_effect_2009}. The working principle is compared directly against that of a standard and split finger IDT, shown in Fig.~\ref{fig:Concept}(d) and (e), respectively. To test the viability of the nanowire transducers as means of launching and receiving a SAW mode, they are configured as an acoustic delay line, as shown in Fig.~\ref{fig:Concept}(f). Here, the blue arrows indicate SAW propagation and serve as the demonstration of this work.

\subsection{Platform Material Selection}
Niobium nitride, a type-II dirty superconductor, is commonly used to fabricate superconducting devices, such as SNSPDs and transition edge sensors (TESs) \cite{venza2025research}. It is a popular thin film superconductor due to its high kinetic inductance, high critical temperature $T_c$, and short coherence length of $ \xi \approx5$ nm. Since reset characteristics are not critical in this application, NbN was implemented primarily for fabrication convenience. The NbN thin film used was deposited with a thickness $d = 15$ nm, a sheet resistance $R_s = 407$  $\Omega /sq$, $T_c \approx 10$ $K$ and a calculated kinetic inductance $L_k \approx 63$ pH/sq.

Scandium aluminum nitride (ScAlN) has emerged as a promising piezoelectric material for acoustic devices, owing to its in-plane isotropic crystal structure and substantially enhanced piezoelectric coefficients relative to its counterpart, aluminum nitride (AlN) \cite{caro_piezoelectric_2015}. The incorporation of scandium into the AlN wurtzite lattice softens the crystal and therefore reduces the acoustic phase velocity. This phase velocity reduction is crucial for enabling high-performing slow-on-fast heterostructure platforms, in which a lower velocity piezoelectric is deposited on a higher velocity substrate. In these platforms, acoustic energy becomes vertically confined for the support of surface acoustic waves (SAWs), namely the fundamental Rayleigh and Sezawa modes~\cite{guida_focused_2024}. Silicon carbide is an ideal substrate for this architecture due to its high acoustic phase velocity~\cite{cho_acoustic_2022}. The large acoustic impedance mismatch between ScAlN and SiC mitigates mode leakage and enables high performing SAW devices. The ScAlN on SiC heterostructure therefore represents an ideal piezoelectric platform \cite{guida_guided_2025}, combining the high electromechanical coupling of ScAlN with the low-loss, high-velocity acoustic properties of SiC, and is selected as the foundation for the transducer demonstrated in this work.

A notable advantage of this platform and technique is the fabrication simplicity. The entire transducer is realized in a single trace nanowire which is patterned in one mask on a planar heterostructure.

%% file: Tex/Modeling.tex
\section{\label{sec:Modeling} Multiphysics Modeling}
The meandered nanowire transducer involves the coupled interaction of three distinct physical domains: the electromagnetic response of the superconducting nanowire, the piezoelectric constitutive behavior of the ScAlN, and the elastic dynamics of the ScAlN on SiC heterostructure. COMSOL Multiphysics is utilized for coupled electromagnetic, piezoelectric, and solid mechanics physics. The goal of the simulation is to demonstrate electromechanical coupling in a meandered nanowire transducer, where the periodic geometry of the nanowire drives acoustic modes through piezoelectric coupling, and this framework provides direct numerical validation of that working principle. Here, the acoustic wavelength and aperture are set to $\lambda_A$ = 1.4 $\mu$m and $f_{EM,o}$ = 120 GHz, respectively, to balance mesh refinement with computational efficiency, though the underlying physics and working principle are independent of these specific values.

The simulation begins with a full-wave frequency-domain simulation of the meandered NbN nanowire. The nanowire is defined with a thickness of $t = 15$ nm and electrical conductivity $\sigma = \frac{1}{j\omega L_k t}$, where $\omega$, $L_k$, and $t$ represent the electromagnetic angular frequency, kinetic inductance, and thickness of the NbN, respectively. This conductivity captures the dominant inductive response of the superconducting film below $T_c$, yielding a purely imaginary value consistent with the lossless and reactive behavior of superconducting films at microwave frequencies. The geometry is implemented such that the meandered nanowire lies in the $xy$ plane. Input and output lumped ports are defined at the terminals of the nanowire, allowing the solver to compute the spatially resolved electromagnetic field distribution throughout the device at the electromagnetic resonant frequency, $f_{EM,o}$, which is defined by the meander aperture and effective electromagnetic wavelength $\lambda_{EM}$, while the meander pitch defines the acoustic wavelength, $\lambda_A$. From the generalized Amp\`{e}re's Law, the time-harmonic current distribution in the nanowire is written as:

\begin{equation}
    I = \iint_S j\omega \, \mathbf{D} \cdot \hat{\textbf{n}} \, dA
    \label{eq:Current}
\end{equation}
\input{Figs_Tex/COMSOL}
Here, the current distribution ($I$) is solved for by integrating the normal component of the electric displacement field ($\textbf{D}$) over the nanowire surface ($S$), where $\hat{\textbf{n}}$ is the outward surface normal and $\omega$ is the angular frequency tone injected into the nanowire. While the electromagnetic drive governs the excitation of the acoustic mode, the full piezoelectric and elastic response of the heterostructure must be modeled to capture the electromechanical coupling and resulting acoustic wave dynamics. To achieve this, the electromagnetic displacement field solved at $f_{EM,o}$ is identity-mapped onto the same geometry, serving as the excitation source for the coupled piezoelectric and solid mechanics physics. In COMSOL, identity mapping copies a solution from one boundary or domain and applies it to another. The mapped electric displacement field acts as an effective drive that transfers the spatially periodic electromagnetic excitation directly to the piezoelectric layer. Floquet periodic boundary conditions are applied along the boundaries normal to the $x$-direction to simulate an infinite periodic transducer within a unit cell model, and a ground plane is defined at the bottom of the SiC substrate consistent with the experimental device configuration. Since the nanowire is continuous, geometric partitions are introduced to define terminal regions analogous to electrode definitions in a standard IDT, enabling extraction of an effective admittance response.

With the identity-mapped fields applied as boundary conditions, a frequency-domain sweep over the acoustic frequency range is performed to capture the acoustic response of the platform. The admittance is defined as $Y_{11} = I/V$, where the current $I$ is evaluated using Eq.~\ref{eq:Current} and the voltage $V$ is calculated as the difference between an average operator applied to two consecutive antiparallel nanowire segments. This difference is necessary because the true ground reference is placed at the bottom of the substrate to mimic the experimental implementation. Across one full period of the field distribution, the effective drive voltage is thus calculated, yielding a physically meaningful admittance response.

For direct comparison, a standard IDT counterpart is built and simulated using the same material stack with 30~nm aluminum electrodes. The same boundary conditions apply, with the only difference lying in the method of exciting the transducer conducting material. The $z$-component of the electric displacement field, which is directly proportional to the current, is compared between the standard IDT and the meandered nanowire in Fig.~\ref{fig:COMSOL}(a) and (b), respectively at each of their acoustic resonant frequencies. This comparison highlights the alternating-polarity electromagnetic fields that are analogous to those of a standard IDT, constituting proof of the working mechanism of the meandered superconducting nanowire as an acoustic wave transducer.
\input{Figs_Tex/ModesDisp}
The resulting admittance spectra for both the standard IDT and the meandered nanowire are shown in Fig.~\ref{fig:COMSOL}(c). The spectrum exhibits the characteristic resonance and antiresonance feature consistent with a modified Butterworth-Van Dyke (mBVD) response, confirming electromechanical coupling between the nanowire excitation and the acoustic modes of the ScAlN on SiC stack. The differences in the acoustic resonance ($f_{A,o}$) likely come from the difference in transducer material and thickness, also impacting the transducer static capacitance \cite{colombo_sub-ghz_2022}. 

A key property of SAW modes is that their displacement is dominated by the out-of-plane component. The resulting $z$-directed displacement at $f_{A,o}$ for a drive at $f_{EM,o}$, shown in Fig.~\ref{fig:Modes}(a), demonstrates the successful transduction of the fundamental Rayleigh SAW mode. Additionally, the identity mapping is repeated for precisely an electromagnetic drive frequency of 2$f_{EM,o}$. The out-of-plane displacement at the same acoustic frequency, $f_{A,o}$, is shown in Fig.~\ref{fig:Modes}(b), demonstrating the ability to launch a higher-order Rayleigh mode.

Collectively, these simulation results confirm that the meandered nanowire transducer achieves electromechanical coupling through a mechanism that is governed by two independently tunable geometric parameters: one controlling the electromagnetic resonance and one controlling the acoustic resonance. This decoupling is the key feature of the transducer architecture, and the numerical modeling serves as a powerful tool to visualize the working principle prior to experimental validation.

%% file: Figs_Tex/COMSOL.tex
\begin{figure}[b!]
\centering
\includegraphics[width=\linewidth]{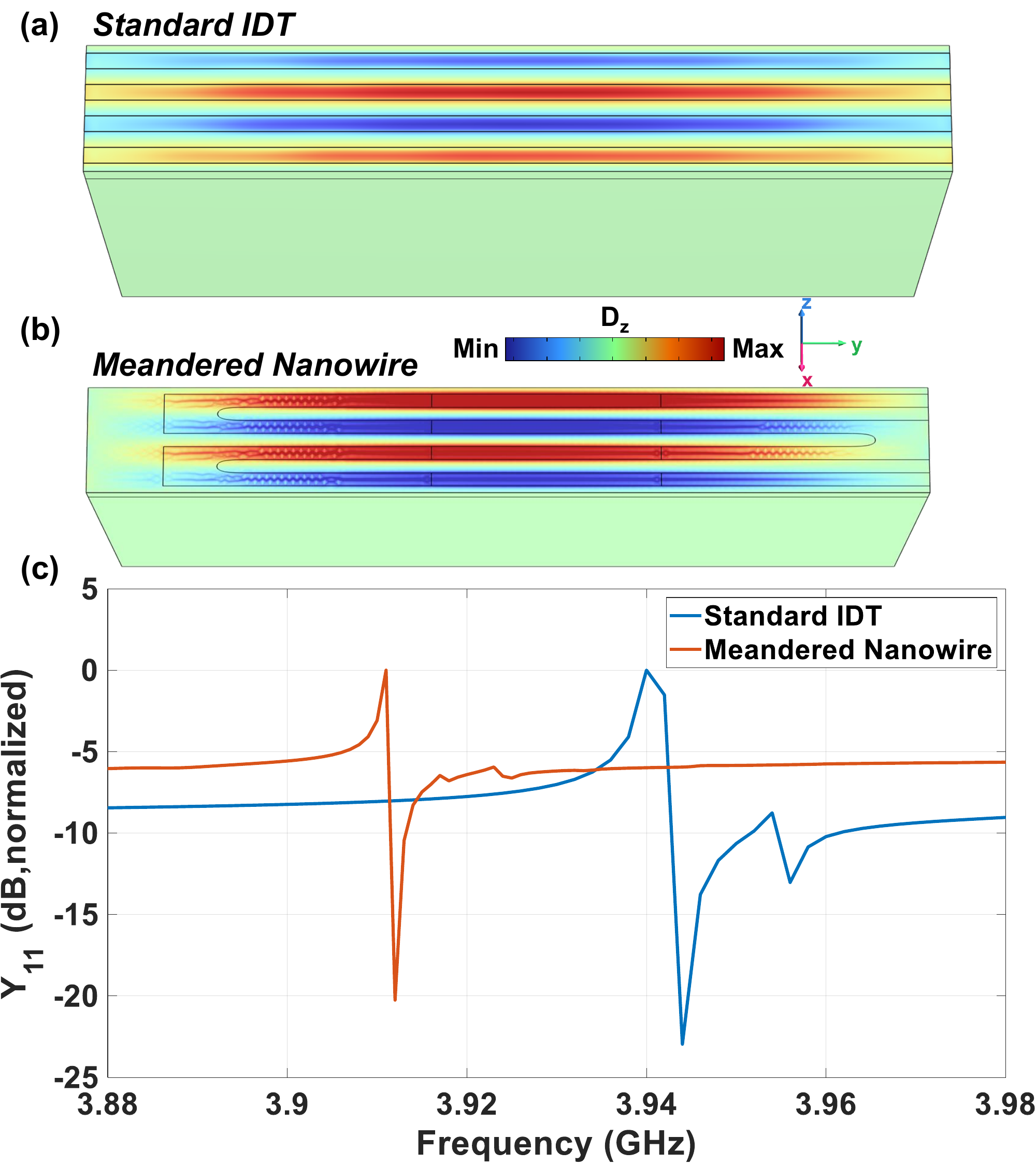}
\caption{The $z$-component of the electric displacement field, $D_z$, proportional to current, for (a) a standard IDT at resonance and (b) the meandered nanowire transducer with equivalent apertures and acoustic wavelength driven at $f_{EM,o}$. (c) Normalized admittance spectra for the standard IDT and the meandered nanowire transducer, each exhibiting the characteristic resonance and anti-resonance of an electromechanically coupled transducer.}
\label{fig:COMSOL}
\end{figure}

%% file: Figs_Tex/ModesDisp.tex
\begin{figure}[b!]
\centering
\includegraphics[width=\linewidth]{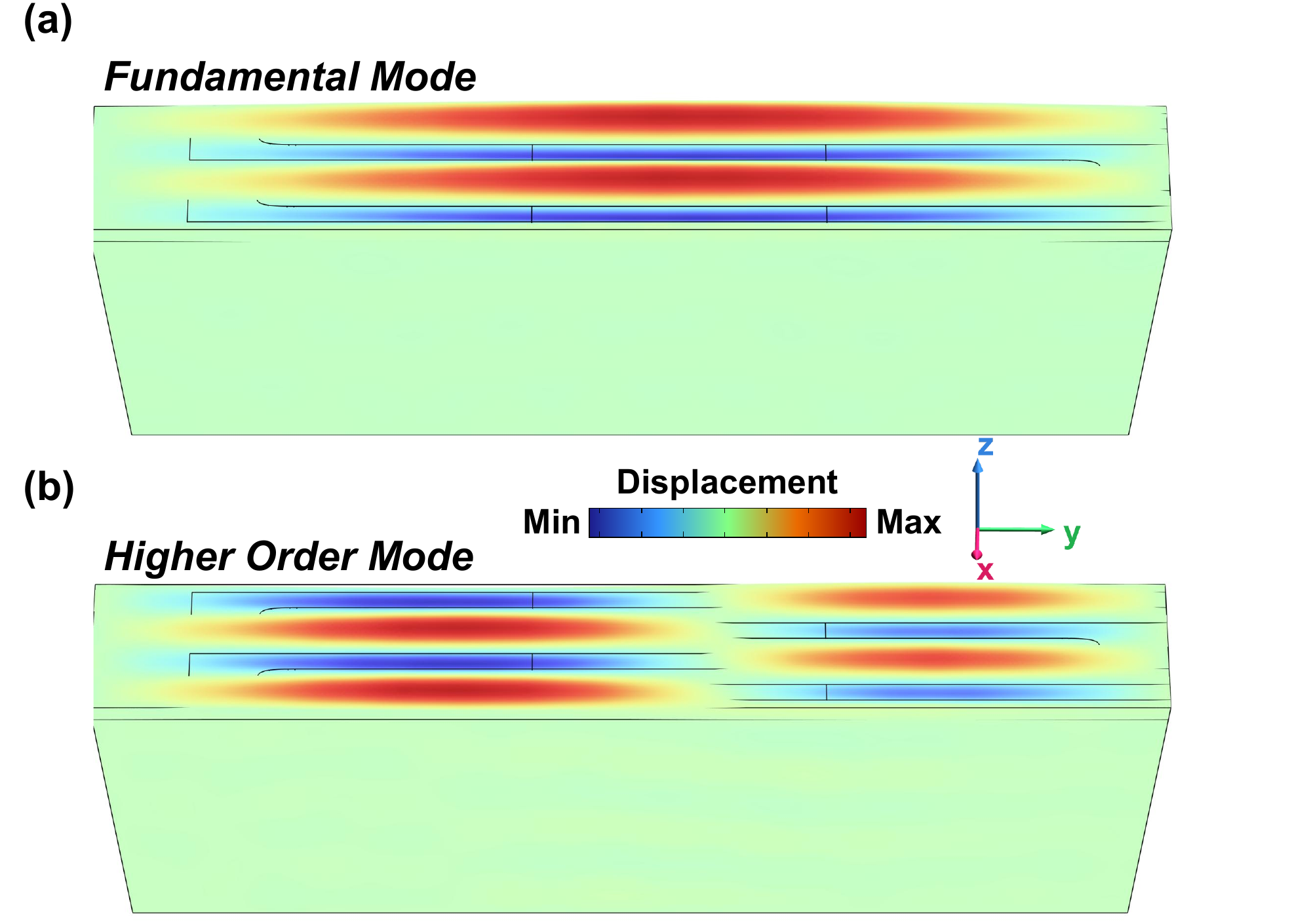}
\caption{Out-of-plane displacement of the meandered nanowire transducer driven at (a) $f_{EM,o}$, exciting the fundamental Rayleigh SAW mode, and (b) 2$f_{EM,o}$, exciting the first higher-order mode.}
\label{fig:Modes}
\end{figure}

%% file: Tex/Fabrication.tex
\section{\label{sec:Fabrication and Test Setup} Device Fabrication}
Fabrication of the devices begins with the deposition of a 500 nm layer of 30\% ScAlN using RF magnetron sputtering on a silicon carbide substrate. Following the piezoelectric film growth, 15 nm of NbN is deposited using DC magnetron sputtering. Devices are subsequently patterned with electron-beam lithography using ma-N 2400 negative photoresist and developed with AZ726 developer. Finally, a $SF_6$-based ICP dry etch is used to define the nanowire transducers.  
\input{Figs_Tex/Fab}
The two nanowire transducers are configured in a delay line configuration, with the delay line length (defined as center-to-center distance between transducers) set to 500 $\mu$m. The fabricated delay line is seen through a scanning electron microscope (SEM) image in Fig.~\ref{fig:IVSEM}(a). The nanowire aperture is fabricated with a length of 640 $\mu$m based on an effective permittivity ($\epsilon_{eff}$) of approximately 550, extracted from Sonnet RF simulations. This meandered nanowire is highlighted in Fig.~\ref{fig:IVSEM}(b). The transducers are designed with impedance matched, meandered Klopfenstein tapers \cite{klopfenstein_transmission_1956, zhao_single-photon_2017}. These tapers match the high characteristic impedance of the nanowire meander to 50 $\Omega$, significantly increasing the signal-to-noise ratio (SNR) of the devices \cite{zhu2019superconducting}, and are numerically verified in Sonnet RF simulations.

Following device fabrication, a diced chip containing the device under test (DUT) is mounted to a custom cryogenic PCB using GE-Varnish and wirebonded using aluminum wire to 50 $\Omega$ breakout traces. The PCB is then mounted in a shielded enclosure, allowing cryogenic connections to the devices via SMA cables and brought outside the cryostat. The enclosure was mounted to the 1 K stage of a high performance evaporative cryostat, calibrated and then closed, pumped down, cooled down, and tested at 0.9 K. At this cryogenic temperature below the critical temperature, the superconducting nanowire is expected to have no electrical resistance. Each nanowire transducer is tested individually through a current-swept IV characterization using a source-measurement unit. The results shown in Fig.~\ref{fig:IVSEM}(c) indicate zero voltage response across an input current of -10 to 10 $\mu$A, indicating superconductivity of both transducers.

%% file: Figs_Tex/Fab.tex
\begin{figure}[b!]
\centering
\includegraphics[width=\linewidth]{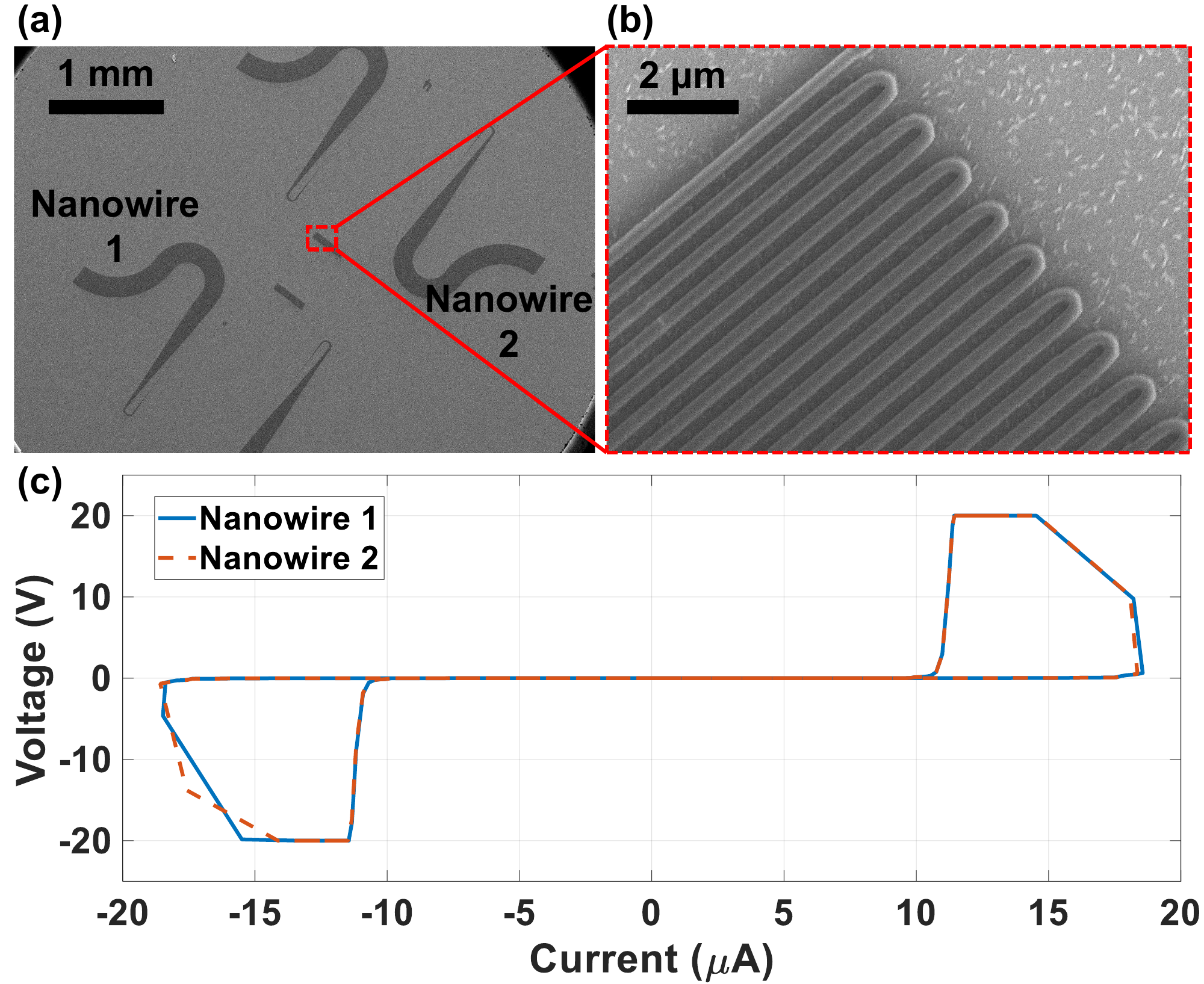}
\caption{(a) Scanning electron microscope (SEM) image of a nanowire-based delay line with impedance matching tapers and (b) single meandered nanowire transducer. (c) IV curve of both nanowire transducers, highlighting superconducting behavior.}
\label{fig:IVSEM}
\end{figure}

%% file: Tex/Results.tex
\section{\label{sec:Results} Experimental Validation}

\input{Figs_Tex/SParams}

The delay line, operating as a microwave transversal filter, is characterized through S-parameter measurements, with the experimental setup shown in Fig.~\ref{fig:SP}(a), with one end of the input and output transducer connected directly to a VNA port, and the other terminated with 50 $\Omega$. The reflection ($S_{11}$) and forward transmission ($S_{21}$) responses are shown in Fig.~\ref{fig:SP}(b) and (c), respectively. Due to calibration drift during the cryostat cooldown, calibration standards were measured and a post-processing correction was applied using the linear transformation described in Appendix A, which accounts for the appearance of $S_{11}$ values exceeding 0 dB. The reflection response exhibits a dip consistent with electromechanical transduction, while the transmission response shows a corresponding local maximum, together confirming that a SAW mode is launched and received across the delay line. The first resonance appears near 3 GHz, with subsequent peaks occurring at approximately 3 GHz periodic intervals, consistent with the excitation of higher-order acoustic modes as predicted by simulation. The fundamental mode, expected near 9 GHz, falls within this harmonic series but cannot be unambiguously identified without spatial imaging of its mode profile.

Acoustic delay lines also act as temporal filters, where the reduced acoustic phase velocity implements a measurable group delay between input and output transducers, governed by $t_g = D/v_g$, where $t_g$, $D$, and $v_g$ represent the group delay, delay line distance, and group velocity, respectively. To extract the time-domain response, an electromagnetic feedthrough gate is applied in the time domain to suppress early-time ringing, followed by a Kaiser window to reduce spectral leakage. An inverse Fourier transform then maps the response into the time domain. The resulting time-domain $S_{21}$ signal is shown in Fig.~\ref{fig:SPTD}. This plot indicates two distinct signal packets, one falling at roughly 85 ns and the other at 125 ns in time. Using the time-of-flight relation with a delay line length of 500 $\mu$m, this corresponds to group velocities of roughly 5900 and 4000 m/s, corresponding to Sezawa and fundamental Rayleigh modes, respectively \cite{guida_focused_2024}. 

\input{Figs_Tex/SPTD}

The identification of both the Sezawa and fundamental Rayleigh modes in the time-domain response constitutes direct experimental evidence of acoustic wave transduction by the meandered superconducting nanowire. The transmission is consistent in the S-parameter characterization and the extracted group velocities are in strong agreement with the well-studied dispersion of the ScAlN on SiC platform. However, since the S-parameter measurements require post-processing due to calibration drift, a second experimental implementation is also conducted, in which the VNA was replaced with a signal generator on the input transducer, and an electrical spectrum analyzer on the output transducer to create a pseudo-$S_{21}$ transmission measurement (in amplitude only). The construction of the measurement, along with the results, are shown in Appendix~\ref{sec:AppB}, which shows excellent agreement with the VNA-measured $S_{21}$ measurement. These results demonstrate, for the first time, that a meandered superconducting nanowire can transduce acoustic waves through piezoelectric coupling, establishing a new class of electromechanical transducer operating at the intersection of superconducting  and acoustic wave microwave devices.

%% file: Figs_Tex/SParams.tex
\begin{figure}[b!]
\centering
\includegraphics[width=\linewidth]{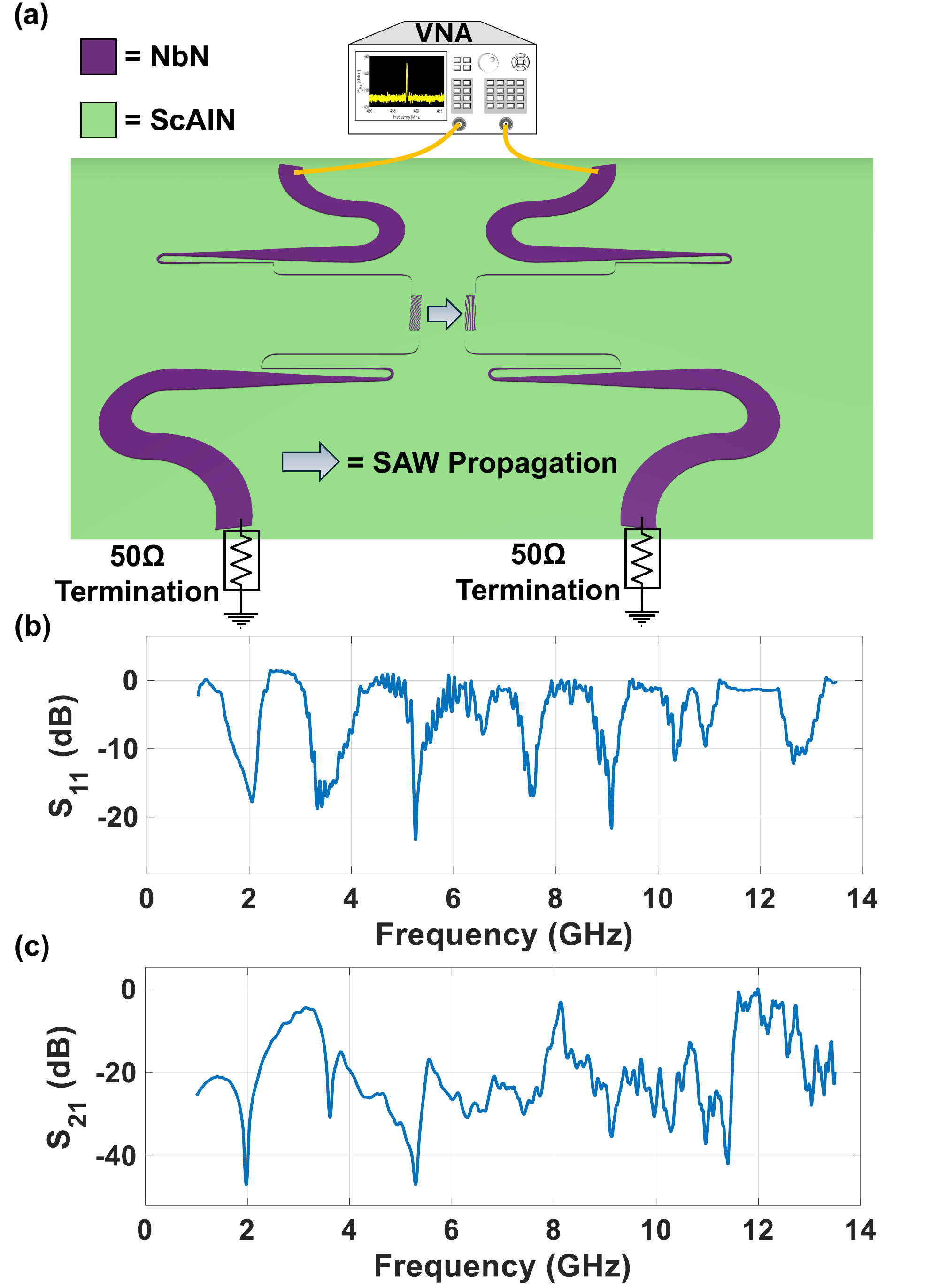}
\caption{(a) Experimental setup with (b) measured $S_{11}$ and (b) $S_{21}$ of the meandered nanowire acoustic delay line.}
\label{fig:SP}
\end{figure}

%% file: Figs_Tex/SPTD.tex
\begin{figure}[b]
\centering
\includegraphics[width=\linewidth]{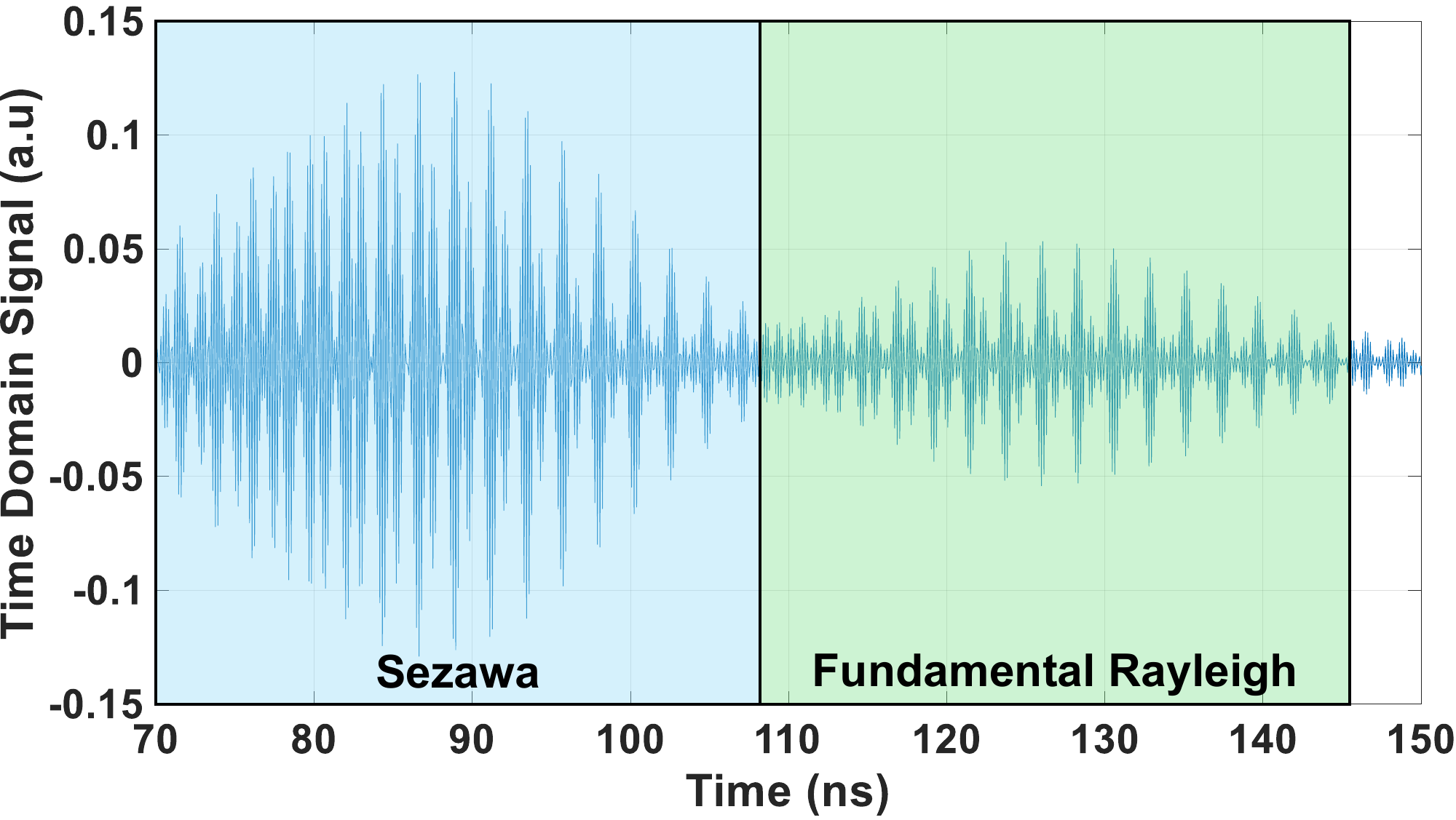}
\caption{Time-domain $S_{21}$ showing temporally distinct signals corresponding to the Sezawa and fundamental Rayleigh SAW modes.}
\label{fig:SPTD}
\end{figure}

%% file: Tex/Conclusion.tex
\section{\label{sec:Conclusion} Conclusion}
In this work, a new class of acoustic wave transducer based on a meandered superconducting nanowire is presented. Unlike standard IDTs, where the operating frequency is lithographically defined by electrode pitch, the meandered nanowire architecture decouples the electromagnetic and acoustic resonances, each governed by independent geometric parameters. This decoupling removes the lithographic upper frequency limit that has historically constrained IDT-based devices, opening a path toward acoustic transduction at frequencies inaccessible to conventional fabrication. This architecture allows the electromagnetic drive frequency to be tuned independently of the acoustic operating point, enabling optimization of insertion loss and electromechanical coupling without sacrificing frequency range. The operating principle is established through a multiphysics simulation framework in COMSOL and validated experimentally through S-parameter characterization and time-domain analysis, where both the fundamental Rayleigh and Sezawa modes are unambiguously identified in time domain analysis.

Future work could include operating the delay line simultaneously as an SNSPD and an acoustic transducer where a single absorbed photon could trigger a lapse in the phonon transmission that can be detected through time-of-arrival measurements. Data could be modulated using photons and detected in the acoustic domain. A more sophisticated multiplexing implementation could be used as an entirely new readout technique for SNSPD arrays.

The implications of this work extend beyond conventional RF signal processing, placing these transducers at an intersection of microwave acoustic devices for quantum applications. This positions the meandered nanowire transducer as a compelling concept for quantum acoustics \cite{chu_quantum_2017}. The demonstration of acoustic transduction through a superconducting nanowire serves as a bridge between acoustic wave devices and technology and the advancing landscape of superconducting quantum technology with implications for both classical and quantum information processing in cryogenic applications.

%% file: Tex/Acknowledgement.tex
\section{\label{sec:Ackowledgement} Acknowledgment}
J.G. acknowledges PhD funding from the NDSEG Fellowship Program. This work was supported  in part by the DARPA Young Faculty Award (YFA) under Grant D24AP00005, the NSF CAREER Program under Award 2340405 and NSF ENG-QUANT under Award 2430923.

%% file: Tex/DataAvailibility.tex
\section{\label{sec:DataAvailability} Data Availability}
The data that supports the findings of this work are not
publicly available. The data is available from the authors
upon reasonable request.

%% file: Tex/AppendixA.tex
\section{Post-Processing Calibration}
Prior to closing the chamber of the cryostat, a calibration was performed with the VNA using an electronic calibration standard. However, over the cooldown time, there was a significant drift in the system calibration, leading to very noisy results. After calibration, measurements on system open, short, and 50 $\Omega$ load were recorded to capture the state of the calibration. 

These measurements were used to perform a post-processing error correction using the one-port, three-term error model \cite{rytting_namodels}. In this mode, errors of the measurement are captured through three error terms, including directivity ($e_{00}$), port match ($e_{11}$), and reflection tracking ($e_{10}e_{01}$). For each standard taken, the measurement equation relating the measured reflection coefficient $\Gamma_{M,i}$ to the known ideal value $\Gamma_i$ can be arranged into the linear form in Eq.~\ref{eq:APEQ1}:

\begin{equation}
\label{eq:APEQ1}
    e_{00} + \Gamma_i \Gamma_{M,i}\, e_{11} - \Gamma_i \Delta e = \Gamma_{M,i}
\end{equation}

Here, $\Delta e = e_{00}e_{11} - e_{10}e_{01}$. Applying this to the short ($\Gamma_S = -1$), open ($\Gamma_O = 1$), and load ($\Gamma_L = 0$) standards yields the 3x3 linear system shown in Eq.~\ref{eq:APEQ2}

\begin{equation}
\label{eq:APEQ2}
    \begin{bmatrix} 1 & -\Gamma_{M,L}\,\Gamma_L & \Gamma_L \\ 
                    1 & -\Gamma_{M,O}\,\Gamma_O & \Gamma_O \\ 
                    1 & -\Gamma_{M,S}\,\Gamma_S & \Gamma_S 
    \end{bmatrix}
    \begin{bmatrix} e_{00} \\ e_{11} \\ e_{10}e_{01} \end{bmatrix}
    =
    \begin{bmatrix} \Gamma_{M,L} \\ \Gamma_{M,O} \\ \Gamma_{M,S} \end{bmatrix}
\end{equation}

This is solved independently at both ports to recover each set of error terms, and the corrected reflection coefficient is then computed from Eq.~\ref{eq:APEQ3}.
\begin{equation}
\label{eq:APEQ3}
    \Gamma = \frac{\Gamma_M - e_{00}}{e_{10}e_{01} + e_{11}(\Gamma_M - e_{00})}
\end{equation}

%% file: Tex/AppendixB.tex
\section{Pseudo-$S_{21}$ Transmission Measurement and Analysis}
\label{sec:AppB}
In addition to the direct measurement of S-parameters, a pseudo-$S_{21}$ transmission spectrum was reconstructed as a heuristic characterization of the device transmission response across the full measurement band. This measurement is termed ``pseudo'' because it recovers only the magnitude of the transmission and does so point-by-point from a series of single-frequency snapshots rather than from a coherent continuous calibrated sweep with phase information. Here, a continuous wave (CW) tone is injected at each frequency step, and the output power is captured independently at each point. Stringing these captured output powers together across all swept frequencies produces a transmission curve that is functionally analogous to $S_{21}$.

A vector network analyzer (VNA) was used solely as a tunable CW source, injecting a single-frequency tone into the input nanowire transducer at each drive frequency. The output of the receiving nanowire transducer was connected to an electrical spectrum analyzer (ESA), which captured the full output power spectrum at each operating point over the range 0.5-20 GHz. One ESA spectrum was recorded per drive frequency across a sweep spanning 1-20 GHz in 0.2 GHz steps at an input power of 14 dBm applied to the launching transducer. 

A key advantage of this measurement architecture is that the ESA captures the complete output spectrum at each drive frequency, making harmonic content naturally visible. Practical sources generate harmonic tones at integer multiples of the intended drive frequency. These harmonics are present in the signal injected into the device, and because they pass through the device and are captured by the ESA alongside the desired tone, they are available for analysis at no additional measurement cost with no modification to the experimental setup. This harmonic content is repurposed as a set of simultaneous independent CW probes of the device transfer function at multiple frequencies within a single measurement. The validity of this interpretation rests on the principle of superposition, where acoustic transducers operate in the linear regime, meaning the device transmission response is a property of the device alone and is independent of how input energy at frequency $f$ was generated. The $n^{th}$ harmonic arriving at the device at frequency $nf_{in}$ therefore interrogates the device identically to a direct CW excitation at $nf_{in}$ (i.e. the second harmonic of a 3 GHz drive probing at 6 GHz is physically identical to a direct 6 GHz excitation). Every harmonic is therefore an additional probe of the transfer function, a direct consequence of linearity and superposition.

At each drive frequency $f_{in}$, the peak output power was extracted at the fundamental frequency and at each harmonic order $h$ ($h = 1, 2, 3, 4, 5$) by searching for the maximum ESA power within a 150 MHz window centered on the expected harmonic frequency $h \times f_{in}$. Harmonics whose expected frequency exceeded the 20 GHz upper bound of the ESA were automatically excluded. The output power for each harmonic was converted and normalized to the peak value observed for that harmonic across the full frequency range, removing the large absolute power differences between harmonic orders that arise from the source.

The challenge in interpreting these harmonic traces is separating true device transmission from noise. Because the VNA source injects harmonic content at every drive frequency, the ESA will register a signal at every harmonic frequency regardless of whether the device is doing anything meaningful at that frequency. A SAW resonance at $f = 4$ GHz, for example, must appear at that frequency regardless of which harmonic order is used to probe it. It will appear on the fourth harmonic when $f_{in} = 1$ GHz, on the second harmonic when $f_{in} = 2$ GHz, and on the fundamental when $f_{in} = 4$ GHz. Noise and measurement artifacts, by contrast, are uncorrelated across harmonic orders and are unlikely to produce coincident peaks at the same absolute frequency across multiple independent traces, and therefore agreement across harmonics is used to identify true transmission.
\input{Figs_Tex/PseudoS21}
Each harmonic has a different native point spacing on the output frequency axis (the fundamental tone has points every 200 MHz, the fundamental step size). The second harmonic has points every 400 MHz, the third every 600 MHz, and so on. All five harmonic curves are first resampled onto a common frequency grid spanning 1-20 GHz before any harmonic comparison is performed. Peaks are then detected in each resampled harmonic curve above a minimum prominence threshold of 0.1 on the normalized 0-1 scale. For each detected peak at output frequency $f$, a Gaussian-shaped vote is cast on the shared frequency axis centered at $f$ with width $\sigma = 0.4$ GHz. The Gaussian spreading is necessary because two harmonics identifying the same resonance may land at slightly different grid points due to their different native spacings, where a hard frequency bin comparison could incorrectly miss these coincidences. Two tallies are then accumulated across all harmonic orders:

\begin{itemize}
    \item \textbf{Vote count:} each harmonic contributes at most 1 to the tally at any frequency. This is enforced by normalizing each harmonic's presence map to a maximum of unity before accumulation, so that a harmonic with many spurious peaks does not dominate the tally over one with a single sharp resonance. A vote count of 3 at $f = 4$ GHz means that $H_1$, $H_2$, and $H_3$ independently identified a peak at that frequency.

    \item \textbf{Weighted prominence:} the same procedure is followed, but instead of contributing a flat $+1$, each harmonic contributes its peak prominence value as the Gaussian height, rewarding sharper and more isolated peaks over weak features that barely exceed the detection threshold. A peak with prominence 0.8 contributes $8\times$ more to the tally than one at the 0.1 minimum threshold.
\end{itemize}

Together these form a heuristic scoring function that amplifies frequency agreement across harmonic orders, with the results shown in Fig.~\ref{fig:PseudoS21}. A high vote count at a given frequency indicates that many independent harmonic probes identified a resonance and a high weighted prominence indicates that those identifications were made with strong peaks. The combination of both (many harmonics agreeing, each with high magnitude) represents the strongest possible evidence within this framework for a true acoustic resonance, and is less likely to arise from noise than any single harmonic observation alone. These scores are  compared against the direct $S_{21}$ measurements to compare how well the heuristic consensus frequencies correspond to features in the closed-loop VNA transmission measurement.

%% file: Figs_Tex/PseudoS21.tex
\begin{figure}[t!]
\centering
\includegraphics[width=\linewidth]{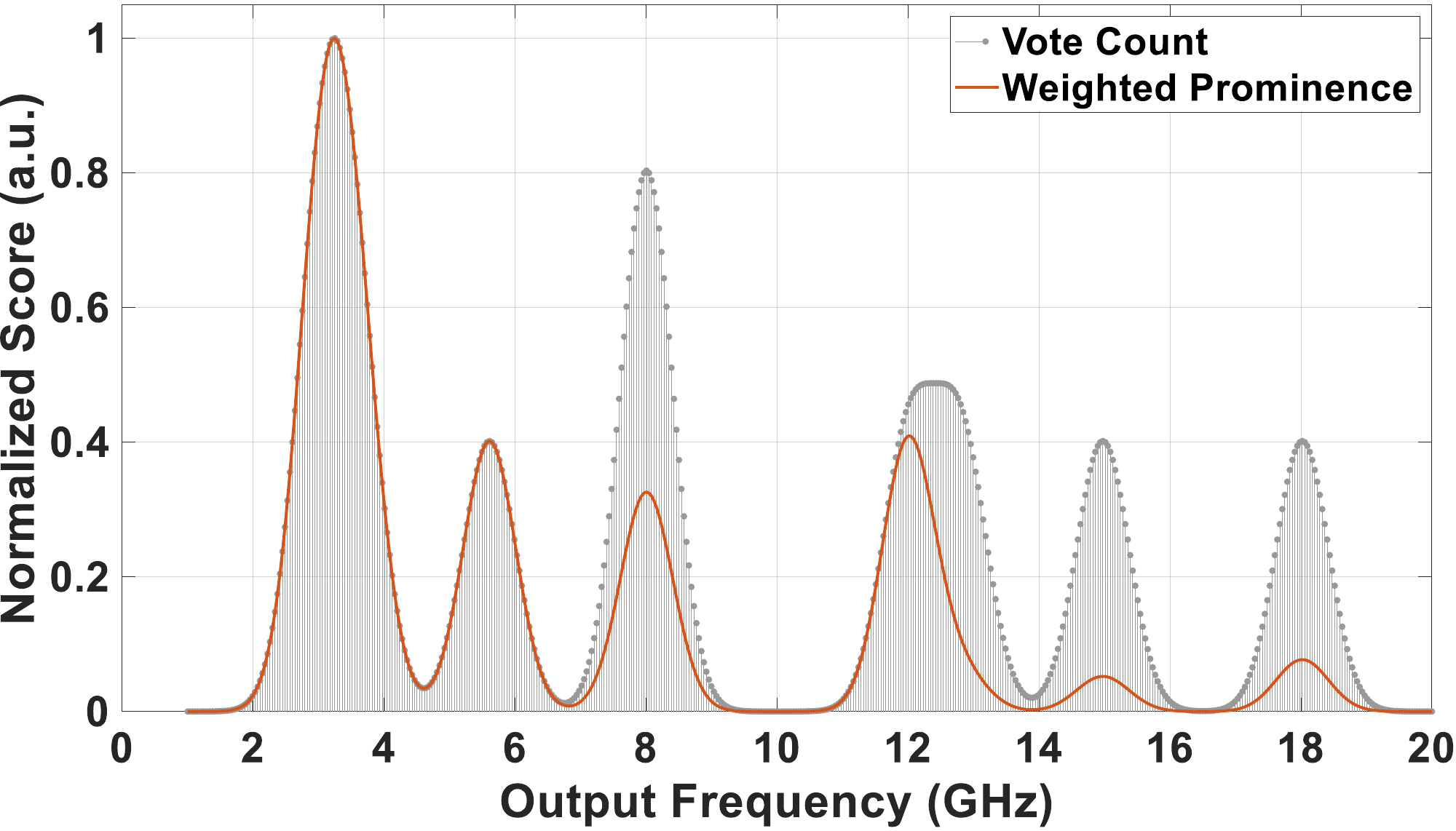}
\caption{Normalized pseudo-$S_{21}$ transmission response reconstructed from electrical spectrum analyzer measurements for a single tone continuous wave sweep. Vote count utilizes the presence of the fundamental tone and its harmonics, and the weighted prominence captures each harmonic's peak prominence, rewarding higher output signals.}
\label{fig:PseudoS21}
\end{figure}